\documentclass[aps,pra,twocolumn,amsmath,amssymb,showpacs,floatfix,superscriptaddress]{revtex4-1}

\usepackage{times}
\usepackage[utf8x]{inputenc}
\usepackage[english]{babel}
\usepackage[T1]{fontenc}
\usepackage{lmodern}
\usepackage{amssymb}
\usepackage{amsmath}
\usepackage{amsthm}
\usepackage[pdftex]{graphicx}
\usepackage{epstopdf}
\usepackage{braket}
\usepackage[bottom]{footmisc}
\usepackage{dcolumn}
\usepackage{bm}
\usepackage{color}
\usepackage{relsize,dsfont,mathrsfs,empheq,verbatim,upgreek}
\usepackage{etoolbox}
\usepackage[caption = false]{subfig}

\usepackage{enumerate}

\newcommand{\Tr}{\mathrm{Tr}}
\newcommand{\Lt}{\mathcal{L}}

\renewcommand{\AA}{\mathcal{A}}

\newcommand{\CC}{\mathcal{C}}

\newcommand{\LL}{\mathcal{L}}

\newcommand{\ii}{\text{i}}
\newcommand{\tr}[2]{\text{Tr}_{ #1 } \left\{ #2 \right\}}

\newcommand{\tS}{\text{S}}

\newcommand{\tB}{\text{B}}
\newcommand{\tSB}{\text{SB}}

\newcommand{\e}{\text{e}}

\renewcommand{\vec}[1]{\mathbf{#1}}

\newcommand{\ph}{{\phantom{\dagger}}}
\renewcommand{\bm}[1]{\textbf{\textit{#1}}}

\begin{document}


\title{Quantum thermodynamics of the spin-boson model using the principle of minimal dissipation}


\author{Salvatore Gatto}
\email{salvatore.gatto@physik.uni-freiburg.de}
\affiliation{Institute of Physics, University of Freiburg, Hermann-Herder-Str. 3, D-79104 Freiburg, Germany}

\author{Alessandra Colla}
\email{alessandra.colla@physik.uni-freiburg.de}
\affiliation{Institute of Physics, University of Freiburg, Hermann-Herder-Str. 3, D-79104 Freiburg, Germany}
\affiliation{Dipartimento di Fisica ``Aldo Pontremoli'', Università degli Studi di Milano, Via Celoria 16, I-20133 Milan, Italy}

\author{Heinz-Peter Breuer}
\affiliation{Institute of Physics, University of Freiburg, Hermann-Herder-Str. 3, D-79104 Freiburg, Germany}
\affiliation{EUCOR Centre for Quantum Science and Quantum Computing, University of Freiburg, Hermann-Herder-Str. 3, D-79104 Freiburg, Germany}

\author{Michael Thoss}
\affiliation{Institute of Physics, University of Freiburg, Hermann-Herder-Str. 3, D-79104 Freiburg, Germany}

\begin{abstract}

A recently developed approach to the thermodynamics of open quantum systems, on the basis of the principle of minimal dissipation, is applied to the spin-boson model. Employing a numerically exact quantum dynamical treatment based on the hierarchical equations of motion (HEOM) method, we investigate the influence of the environment on quantities such as work, heat and entropy production in a range of parameters which go beyond the weak-coupling limit and include both the non-adiabatic and the adiabatic regimes. The results reveal significant differences to the weak-coupling forms of work, heat and entropy production, which are analyzed in some detail.

\end{abstract}

\date{\today}

\maketitle

\section{Introduction}\label{sec:Intro}

The field of quantum thermodynamics has witnessed significant attention and advancement in recent years \cite{Gemmer2004,Schaller2014,Binder2018,Deffner2019,Landi2021}. A central challenge in this area revolves around establishing consistent definitions for work, heat, and entropy production in open quantum systems coupled to thermal reservoirs \cite{Breuer2002,Kosloff2015,Segal2021}. 
While the fundamentals of quantum thermodynamics are well-established for ultra-weak system-bath coupling and continuous information loss,  defining thermodynamic quantities outside of this regime remains a central topic of ongoing research \cite{Weimer2008,Esposito2010,Teifel2011,Alipour2016,Seifert2016,Strasberg2017,Rivas2020,Alipour2021,Landi2021, Tanimura2019}. 

Challenges arise, in particular, for strong system-environment coupling, structured environmental spectral densities, or low temperatures, when non-Markovian dynamics and memory effects become significant, rendering traditional Markovian master equations inadequate \cite{Breuer2016a,deVega2017}. A related issue, which has received considerable attention recently \cite{Campell2018,Strasberg2019,Rivas2020,Alipour2017}, is the connection between non-Markovian dynamics and the emergence of negative entropy production rates. 

Among the different questions considered in the field, two aspects appear particularly important. 
First, at stronger coupling of the open quantum system to its environment, the interaction energy between system and environment becomes non-negligible, requiring an understanding of its contribution to the system's internal energy, work, or heat. Second, the information exchange between the system and its environment, linking information theory and thermodynamics, introduces memory effects and information backflow as crucial aspects in the theoretical treatment \cite{Breuer2016a}.

A variety of approaches have been proposed to address these issues \cite{Weimer2008,Esposito2010,Alipour2016,Strasberg2017,Dou2018,Rivas2020,Thoss2020,Landi2021,Huang2022a,Huang2022b,Zhou2024,Tanimura2022c,Tanimura2022,Tanimura2022b}.
In this work, we consider a recently proposed approach based on the
principle of minimal dissipation \cite{Colla2022a,Picatoste2023}, which leads to a unique decomposition of the quantum master equation into a Hamiltonian and a dissipative part, allowing to identify the contributions describing work and heat. In this approach, thermodynamic quantities are expressed in terms of an effective system Hamiltonian $K_S$ that results from the coupling of the system to its environment. This emergent effective Hamiltonian encapsulates information of the environment, providing insights into how the system responds to the interaction. 

Here, we apply this novel approach to investigate the quantum thermodynamics of the spin-boson model. The spin-boson model, which involves a two-level system (or spin) interacting linearly with a bath of harmonic oscillators, is a paradigmatic model to describe dissipative dynamics of open quantum systems \cite{Weiss1999,ThossWang2008,Leggett1987}. Our study employs the hierarchical equations of motion (HEOM) approach \cite{Tanimura1989,Thoss2021}, which allows a numerically exact simulation of the quantum dynamics of the spin-boson model, and considers a broad parameter range, including both the non-adiabatic and the adiabatic regimes.
To the best of our knowledge, this is the first application of the quantum thermodynamics approach of Ref. \cite{Colla2022a} to a nonintegrable model using a nonperturbative method.
The focus of our investigation is the analysis of the effective Hamiltonian $K_S$, in particular, how the coupling to the thermal bath induces deviations of  $K_S$ from the system Hamiltonian in the different parameter regimes. 
Moreover, we investigate relevant thermodynamic quantities such as work, heat, and entropy production. Our study shows  that the results obtained within the principle of minimal dissipation differ from those  obtained using traditional expressions defined in the weak-coupling limit.

The paper is organized as follows.
In Sec.~\ref{sec:qthermo} we recapitulate the principle of minimal dissipation and introduce all relevant thermodynamics quantities, namely work, heat and entropy production rate. 
The spin-boson model is introduced in Sec.~\ref{sec:micro-model}, together with the HEOM approach.
The results are discussed in Sec.~\ref{sec:results}, focusing on three different parameter regimes of the spin-boson model.
Sec.~\ref{sec:conclu} concludes with a summary. 

\section{Theory}\label{sec:theory}

\subsection{Principle of minimal dissipation}\label{sec:qthermo}

In order to investigate non-equilibrium quantum thermodynamics, we consider an open quantum system $S$ interacting with an environment $B$.  
The Hamiltonian of the composite system is expressed as
\begin{equation} \label{ham-total}
	H(t) = H_S(t) + H_{SB}(t) + H_B ,
\end{equation}

\noindent
where $H_S(t)$ and $H_B$ correspond to the system and environment, respectively, and $H_{SB}(t)$ denotes the interaction. Notably, both the system and interaction Hamiltonians may be time-dependent, capturing phenomena such as external driving or the modulation of the system-environment interaction. The quantum states of $S$ and $B$ are characterized by density matrices $\rho_S$ and $\rho_B$, respectively. The state of the composite system $S+B$ is represented by $\rho_{S+B}$. Assuming a factorized initial state, i.e. $\rho_{S+B}(0)=\rho_{S}(0)\otimes\rho_{B}(0)$, the time-evolution of the quantum state of $S$ defines the dynamical map 
\begin{equation} \label{map}
	\rho_S(t) = \Phi_t\rho_S(0) ,
\end{equation}
where $\rho_S(t)$ is obtained by tracing over the degrees of freedom (DOFs) of the environment of the total density matrix $\rho_{S+B}(t)$,
\begin{align}
	\rho_S(t) = \tr{\tB}{\rho_{S+B}(t)}.
\end{align}

Here, we employ the approach to quantum thermodynamics outlined in Ref.\ \cite{Colla2022a}, which uses a time-convolutionless master equation governing the density matrix $\rho_S(t)$ of the open system $S$,
\begin{equation} \label{tcl-meq}
	\frac{d}{dt}\rho_S(t) = \Lt_t\rho_S(t) = -i \left[K_S(t),\rho_S(t)\right] + {\mathcal{D}}_t  \rho_S(t).
\end{equation}
This exact time-local differential equation involves a time-dependent generator $\Lt_t$, which  captures all memory effects in the open system dynamics \cite{Shibata1977,Shibata1979,Breuer2002}, and incorporates two different contributions. The first is a Hamiltonian contribution, expressed through the commutator with an effective Hamiltonian $K_S(t)$. The second component is a dissipator given by
\begin{equation} \label{dissipator}
	{\mathcal{D}}_t \rho_S = \sum_{k}\theta_{k}(t)\Big[L_{k}(t)\rho_S L_{k}^{\dag}(t) - \frac{1}{2}\big\{L_{k}^{\dag}(t)L_{k}(t),\rho_S\big\}\Big],
\end{equation}
incorporating decay rates $\theta_k(t)$ and time-dependent Lindblad operators $L_k(t)$. 

In the master equation \eqref{tcl-meq}, the splitting of the generator $\Lt_t$ into Hamiltonian and dissipative parts is not unique.
Introducing a \textit{principle of minimal dissipation} \cite{Colla2022a}, it is possible to uniquely define the Hamiltonian part, which determines the Hermitian operator $K_S(t)$ in the commutator as the effective system Hamiltonian, representing the internal energy of the system.
Based on the approach developed in Ref.\ \cite{Hayden2021}, this principle states that the dissipator must be minimal with respect to a certain superoperator norm, a property which is achieved when the Lindblad generators $L_k(t)$ are traceless, leading to a unique expression for $K_S(t)$.

Based on this separation, the internal energy of the system is defined as the expectation value of the effective Hamiltonian,
\begin{equation} \label{internal-energy}
	U_S (t) = \Tr \{K_S(t)\rho_S(t)\}.
\end{equation}
The first law describing the change in internal energy is then formulated as
\begin{equation}\label{first-law}
	\Delta U_S (t) \equiv U_S(t) - U_S(t_0) = W_S(t) + Q_S(t),
\end{equation}
with work $W_S(t)$ and heat $Q_S(t)$ contributions defined as
\begin{align}
	W_S(t) &= \int_{t_0}^t d\tau \, \Tr \big\{ \dot{K}_S(\tau) \rho_S(\tau) \big\},  \label{work} \\
	Q_S(t) &= \int_{t_0}^t d\tau \, \Tr \big\{ K_S(\tau) \dot{\rho}_S(\tau)  \big\}.  \label{heat}
\end{align}
The definition of entropy production is formulated in accordance with the above definition of heat,
\begin{eqnarray}\label{DLutz}
	\Sigma_S (t) = \Delta S_S(t) - \beta Q_S(t) \;,
\end{eqnarray}
where $\Delta S_S(t)=S(\rho_S(t))-S(\rho_S(0))$ is the change of the von Neumann entropy of the reduced system and $\beta=1/\mathrm{k_B}T$ the inverse temperature of the bath. 
Taking the time derivative of Eq.~\eqref{DLutz}, the entropy production rate is obtained,
\begin{eqnarray} \label{entr-prod-rate}
	\sigma_S(t) &\equiv& \dot{\Sigma}_S(t) .
\end{eqnarray}

For weak system-bath coupling, $K_S$ is expected to approach the system Hamiltonian $H_S$, and a weak-coupling version of the thermodynamical observables can be defined accordingly as \cite{Spohn1978,Lebowitz1978,Alicki1979,Kosloff1984,Kosloff2013,Deffner2008Feb,Alicki2018}
\begin{align}
	W_w(t) &= \int_{t_0}^t d\tau \, \Tr \big\{ \dot{H}_S(\tau) \rho_S(\tau) \big\},  \label{workweak} \\
	Q_w(t) &= \int_{t_0}^t d\tau \, \Tr \big\{ H_S(\tau) \dot{\rho}_S(\tau)  \big\}, \\
	\sigma_w(t) &= \partial_t \left(\Delta S_S(t) - \beta Q_w(t)\right).  \label{heatweak}
\end{align}

\subsection{Spin-boson model and HEOM method}\label{sec:micro-model}

As an example of an open quantum system, we consider the spin-boson model \cite{Weiss1999,ThossWang2008} described by 
the Hamiltonian
\begin{align}
	{H} = {H}_\tS + {H}_\tSB + {H}_\tB.
	\label{eq:general_Hamiltonian}
\end{align}
The system Hamiltonian ${H}_\tS$ describes a two-level system (representing, e.g., a spin degree of freedom or electronic states of an electron transfer reaction),
\begin{align}
	{H}_\tS = -\varepsilon \sigma_z +\Delta \sigma_x \, ,
	\label{eq:Hamiltonian:system}
\end{align}
characterized by the energy bias $\varepsilon$ and the coupling $\Delta$ of the two states. Pauli matrices are expressed in the eigenbasis $\{\ket{0},\ket{1}\}$ of $\sigma_z$, i.e. 
\begin{align}
	\sigma_x=\ket{0}\bra{1}+\ket{1}\bra{0} , \,  \sigma_z=\ket{0}\bra{0}-\ket{1}\bra{1} \, .
	\label{eq:Pauli}
\end{align}
The environment consists of a bath of 
harmonic oscillators,
\begin{align}
	{H}_\tB = \sum_{k} \omega^\ph_{k} {b}_{k}^\dagger {b}_{k}^\ph \, ,
	\label{eq:Hamiltonian:bosonic_bath}
\end{align}
interacting with the system through system-environment coupling terms, which are linear in the bath creation/annihilation operators,
	\begin{align}
		{H}_\tSB  =& \sigma_z \sum_{k} \lambda^\ph_{k} \left(  {b}_{k}^\dagger + {b}_{k}^\ph\right) \, .
		\label{eq:Hamiltonian:bosonic_coupling}
	\end{align}
Here, $b^\dagger_{k}$/$b^\ph_{k}$ denote the bosonic creation/annihilation operator associated with mode $k$ of the bosonic bath with frequency $\omega_{k}$. 
The influence of the bosonic environment on the system dynamics is encoded in its spectral density function,
	\begin{align}
		J_{\tB}(\omega) =& \pi \sum\limits_{k} |\lambda^\ph_{k}|^2 \delta(\omega-\omega^\ph_{k}) .
		\label{eq:bosonic_spectral_density_general_definition}
	\end{align}
Throughout this paper we consider a  spectral density in Debye form,
 	\begin{align}
 	J_{\tB}(\omega) =& \frac{\pi}{2}\alpha \, \omega \frac{\omega_c^2}{\omega^2 + \omega_c^2} \, ,
 	\label{eq:debye_spectral_density}
 \end{align}
where $\omega_c$ is the characteristic bath frequency and $\alpha$ is the dimensionless Kondo parameter describing the system-bath coupling strength. 

The spin-boson model is a paradigmatic model to describe dissipative quantum dynamics \cite{Weiss1999,Leggett1987}. Despite its simple form, it has applications to a variety of different processes and phenomena, including electron transfer \cite{Marcus1985} and macroscopic quantum coherence \cite{Weiss1987}. The spin-boson model is also interesting
from a more fundamental point of view as it shows
a transition from coherent dynamics to incoherent decay as
well as a quantum phase transition \cite{Bray1982,Chakra1982,Wang2019}.

In order to investigate the thermodynamic characteristics of the spin-boson model, we use the numerically exact hierarchical equations of motion (HEOM)  approach \cite{Tanimura1989,Schinabeck2016,Thoss2021}.
The central quantity of the approach is the reduced density matrix $\rho_S (t)$ of the subsystem. 
It is assumed that the initial state factorizes, i.e. $\rho_{S+B}(0)=\rho_{S}(0)\otimes\rho_{B}(0)$, with
\begin{align} \label{gibbs-hb}
	\rho_B(0)= e^{- \beta H_B }/Z_B .
\end{align} 
Exploiting the Gaussian properties of the non-interacting environment, a formally exact path-integral for the reduced density matrix of the subsystem can be derived
involving the Feynman-Vernon influence functional \cite{Feynman1951,Feynman1963,Tanimura1989,Jin2008}.

As a consequence of a noninteracting bath with Gaussian statistics, all information about system-bath coupling relevant for the system dynamics is encoded in the time-correlation function of the isolated bath, directly related to the bath spectral density (cf. Eqs. \eqref{eq:debye_spectral_density}) via 
	\begin{align}\label{eq:corr}
		C_{\tB}(t) = & \sum_k |\lambda^\ph_{k}|^2 \tr{\tB}{\left(b_k^{\dag}b_k {\rm e}^{i\omega_k t}+b_k b_k^{\dag}{\rm e}^{-i\omega_k t}\right)\rho_{B}(0)} \nonumber\\
		=&\int_{-\infty}^\infty d\omega\frac{J_{\tB}(\omega)}{\pi} f_B(\omega){\rm e}^{-i\omega t} \, ,
	\end{align}
where $f_B(\omega)=(1-{\rm e}^{-\beta\omega})^{-1}$ is the Bose distribution function.
Within the HEOM approach, the correlation function is expanded as a series of exponential function,
	\begin{align}
		C_{\tB}(t) \approx 
		\sum_{l=0}^{l_{max}} \eta_{l} \e^{-\gamma_{l} t},
	\end{align}
where the weights $\eta_l$ and decay rates $\gamma_l$ are determined by decomposing the Bose distribution function according to the  Pad\'e \cite{Hu2010,Hu2011,Kato2015,Erpenbeck2019,Schinabeck2020}  decomposition scheme and performing the integration in Eq.\eqref{eq:corr} using the theorem of residues. Thereby, lower temperatures require a larger number of poles to properly represent the bath correlation function. For the results considered in this work ($\beta \omega_c=5$, $\beta \omega_c=25$), the decomposition of the bath-correlation functions converges at $l_{max}\le10$.

The hierarchical equations of motion are constructed by consecutive time-derivatives of the influence functional, leading to \cite{Tanimura1989,Schinabeck2018},
\begin{alignat}{2}
	\frac{\partial}{\partial t} \rho^{(n)}_{\bm{l}}(t) =& -\left( i \LL_\tS + \sum_{l=1}^n \gamma_{l} \right) \rho^{(n)}_{\bm{l}}(t)
	\nonumber \\
	&- \sum_{l_x} \AA \rho^{(n+1)}_{\bm{l}^+_x}(t) -\sum_{l=1}^n \CC_{ l}  \rho^{(n-1)}_{\bm{l}^-_l}(t) ,
	\label{eq:general_HQME}
\end{alignat}
with the notation for the multi-index vectors ${\bm{l}} = {l_1 {\cdot}{\cdot}{\cdot}l_n}$, ${\bm{l}^+_x}= {l_1 {\cdot}{\cdot}{\cdot}l_n l_x}$, and ${\bm{l}^-_l} = {l_1 {\cdot}{\cdot}{\cdot}l_{l-1}l_{l+1}{\cdot}{\cdot}{\cdot}l_n}$, and the definition $\LL_\tS O = [H_\tS,O]$.
Here, $\rho^{(0)}(t)$ represents the reduced density operator of the system $\rho_S(t)$, and the higher-tier auxiliary density operators (ADOs) $\rho^{(n)}_{\bm{l}}(t)$ encode the influence of the environment on the subsystem dynamics. Tier by tier the ADOs introduce the effect of higher order system-bath correlations.  These ADOs also include the information on time-local transport observables like energy or particle currents \cite{Jin2008,Haertle2013a,Kato2015,Kato2016}.  
The operators $\AA_{l}$ and $\CC_{l}$ connect the $n$th-tier ADO to the ($n{+}1$)th- resp. $(n{-}1)$th-tier ADOs via
\begin{subequations}  
	\begin{align}
		\AA \rho^{(n)}_{\bm{l}}=&  \left\lbrace \sigma_z   , \rho^{(n)}_{\bm{l}} \right\rbrace_+ \, ,\\
		\CC_{l} \rho^{(n)}_{\bm{l}}=&\, \eta_{l}   \sigma_z  \rho^{(n)}_{\bm{l}} - \eta^*_{l}  \rho^{(n)}_{\bm{l}} \sigma_z,
	    \label{eq:general_HQME_upbuilding_operators}
	\end{align}
\end{subequations}  
leading to a hierarchy of equations of motion. In the numerical calculations, the hierarchy needs to be truncated in a suitable manner. As shown in Appendix \ref{app-A}, for the parameters considered in this work, using up to ten tiers in the hierarchy provides converged results.

The HEOM approach is used below to analyze the effective Hamiltonian $K_S$ of the spin-boson model as well as thermodynamic quantities such as work, heat and entropy. To this end, we make use of the pseudo-Kraus 
representation for the generator, which states that the generator of any Hermiticity and trace preserving map can be written as \cite{Choi1975}
\begin{equation}\label{eq:krausform}
\Lt_t \rho_S(t) = \sum_k \theta_k(t) E_k(t) \rho_S(t) E_k^{\dagger}(t) \, ,
\end{equation}
 with some operators $E_k(t)$ and real (not necessarily
positive) coefficients $\theta_k$(t) fulfilling the condition  
\begin{equation}
	\sum_k \theta_k(t) E_k^{\dagger}(t)E_k(t) = 0 \; .
\end{equation}

To evaluate the pseudo-Kraus representation for $\Lt_t$, we simulate the dynamics for different orthogonal initial states of the system, in order to get a representation of the dynamical map $\Phi_t$. The pseudo-Kraus operators are then obtained by diagonalizing the Choi matrix associated to $\Lt_t$ \cite{Andersson2007}. Additional details are provided in Appendix \ref{app-B}. Following the derivation from Ref.\ \cite{Colla2022a}, the final expression for the effective Hamiltonian reads
\begin{align}
	K_S(t) = -\frac{\ii}{4} \sum_k \theta_k(t) \Big[ &\Tr \{ E_k(t) \} E_k^{\dagger}(t) \nonumber \\
	- &\Tr\{ E_k^{\dagger}(t)\} E_k(t) \Big] \; , 
\end{align}
and the various thermodynamic quantities can be evaluated through numerical integration.

\section{Results}
\label{sec:results}
In this section, we use the concepts introduced above to analyze the spin-boson system for three distinct scenarios: the non-adiabatic unbiased case ($\varepsilon = 0$, $\Delta / \omega_c < 1$), the non-adiabatic biased case ($\varepsilon \neq 0$, $\Delta / \omega_c < 1$), and the adiabatic biased case ($\varepsilon \neq 0$, $\Delta / \omega_c > 1$).  We will consider regimes of moderate to strong system-bath coupling strength $\alpha$, for which the dynamical map, underlying the time-convolutionsless master equation in Eq.\ (\ref{tcl-meq}), is invertible, such that the generator in Eq.\ (\ref{tcl-meq}) is well defined. The primary focus of our investigation is the analysis of the effective Hamiltonian $K_S(t)$, in particular, how the coupling to the thermal bath induces deviations of  $K_S(t)$ from the system Hamiltonian $H_S$ in the different parameter regimes. 
Moreover, we investigate relevant thermodynamic quantities such as the work $W_S(t)$, the heat $Q_S(t)$, and the entropy product rate,  $\sigma_S(t)$, which are compared to the corresponding expressions defined in the weak-coupling limit, $Q_w(t)$ and $\sigma_w(t)$.

\subsection{Unbiased ($\varepsilon=0$), nonadiabatic ($\Delta/\omega_c<1$) regime}

\begin{figure}[t]
	\includegraphics[width=0.50\textwidth]{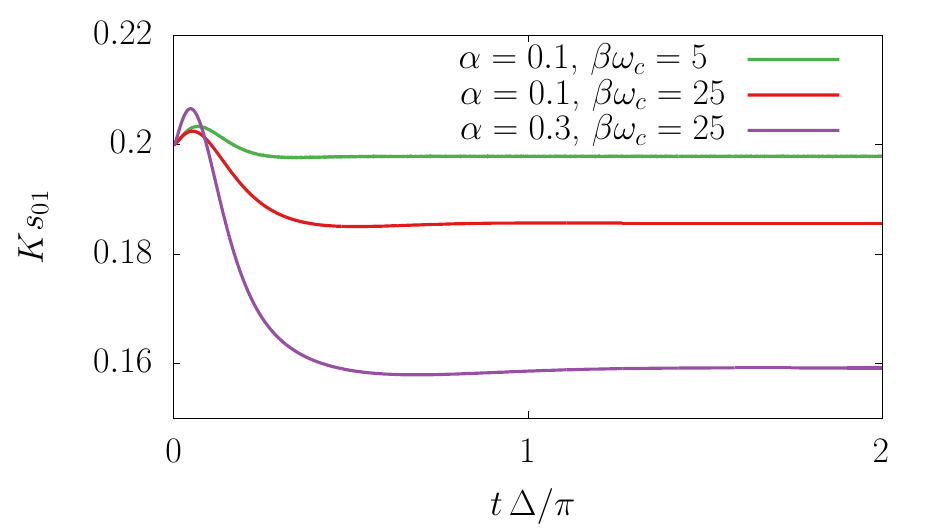}
		\caption{Off-diagonal element of $K_S(t)$ as a function of time for different values of system-bath coupling $\alpha$ and inverse temperature $\beta$. Other parameters are $\varepsilon=0$, $\Delta/\omega_c=0.2$. Results are given in units of $\omega_c$.
	}
	\label{fig:kse0}
\end{figure}
\begin{figure}[t]
	\includegraphics[width=0.50\textwidth]{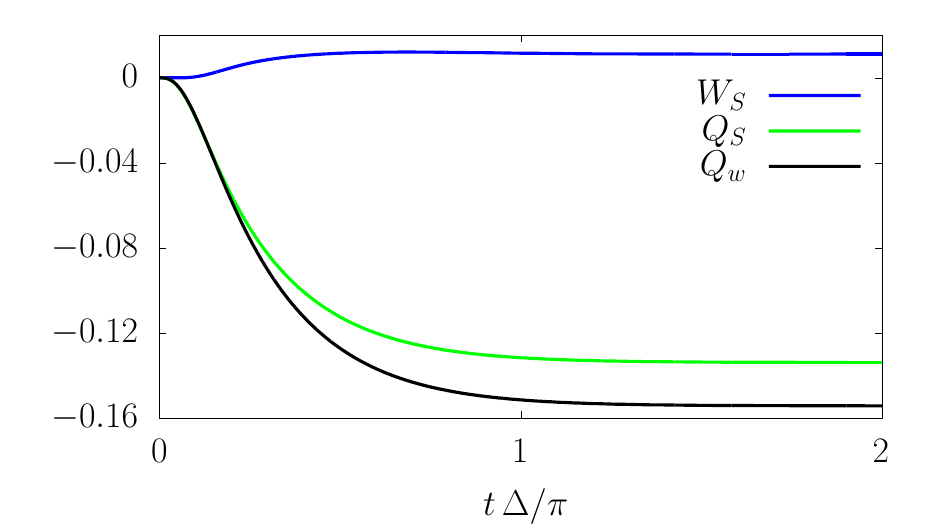}
	\caption{Work $W_S(t)$ and heat $Q_S(t)$ as a function of time, as well as a comparison with the weak-coupling heat $Q_w(t)$. Parameters are $\alpha=0.3$, $\beta \omega_c=25$, $\varepsilon=0$, $\Delta/\omega_c=0.2$. Results are given in units of $\omega_c$.
}
	\label{fig:qwe0}
\end{figure}
\begin{figure}[t]
	\includegraphics[width=0.50\textwidth]{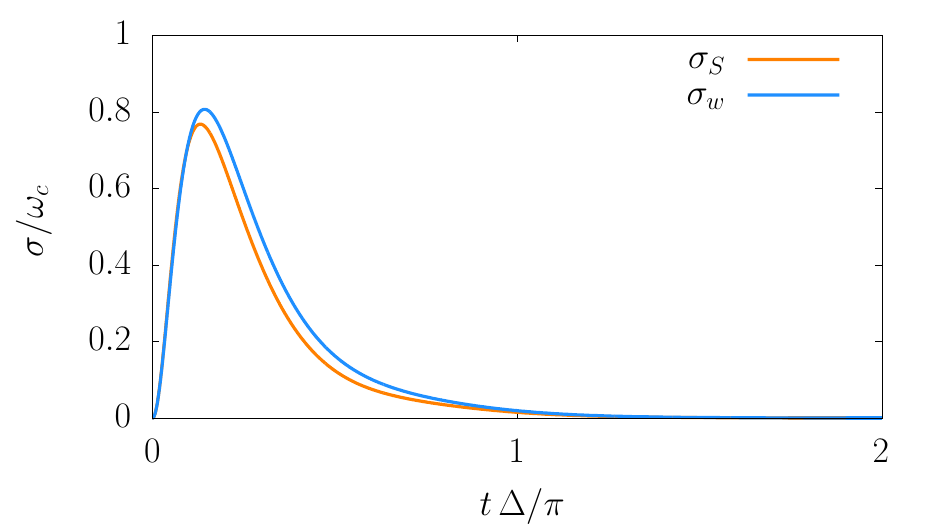}
	\caption{Comparison between the entropy production rate $\sigma_S(t)$ and its weak-coupling version $\sigma_w(t)$. Parameters are $\alpha=0.3$, $\beta \omega_c=25$, $\varepsilon=0$, $\Delta/\omega_c=0.2$.
	}
	\label{fig:sswe0}
\end{figure}

We begin our discussion with the unbiased case, $\epsilon =0$. It is known \cite{Weiss1999,ThossWang2008} that in the non-adiabatic regime ($\Delta/\omega_c<1$) with weak system-bath coupling ($\alpha<0.5$), the spin-boson model exhibits coherent decay of $\langle \sigma_z\rangle (t)$ to its steady state value of zero, while $\langle \sigma_x\rangle (t)$ monotonically decays to some negative value.  In this regime, numerical evidence indicates that the effective Hamiltonian $K_S(t)$ takes 
on a similar form as the system Hamiltonian $H_S$,
\begin{eqnarray} \label{kse0}
	K_S(t) = (\Delta+\delta_{x}(t)) \sigma_x .
\end{eqnarray}  
Thus, for $\varepsilon=0$, the system-bath coupling induces a time-dependent shift, $\delta_{x}(t)$, of the off-diagonal elements of the effective Hamiltonian ${K_S}_{jl}(t)=\bra{j} K_S(t) \ket{l}$, with $j\neq l$.
Notably, since there is no shift in the diagonal elements of $K_S(t)$, the eigenstates of the effective Hamiltonian are time-independent. This finding is a consequence of the unbiased regime: since the two states of the system are degenerate in energy, they interact with the environment in the same way. Thus, the system-bath coupling would induce the same shift on the energy levels, and the system stays unbiased.
Fig.\ \ref{fig:kse0} depicts the off-diagonal elements of the effective Hamiltonian $K_S(t)$ for different parameters of the system-bath coupling and the temperature.
The shift exhibits damped oscillation and approaches a non-zero value for long times. 
This time-dependent modulation of the effective Hamiltonian corresponds to a renormalization of the energy levels of $H_S$ induced by the coupling to the bath. In the considered parameter regime, the shift increases upon increasing system-bath coupling strength and decreasing temperature.

Fig.\ \ref{fig:qwe0} shows the time evolution of work, $W_S(t)$, and heat, $Q_S(t)$, compared to its weak-coupling form $Q_w(t)$. From the parameter sets shown in Fig.  \ref{fig:kse0}, we have selected the most interesting case of larger coupling ($\alpha=0.3$) and low temperature ($\beta \omega_c=25$).  The effects are very similar but less pronounced for smaller coupling and higher temperature.  Since the system Hamiltonian $H_S$ is constant in time, the weak-coupling form of the work, $W_w(t)$, vanishes. As a results of the time-dependent shift in the effective Hamiltonian, $K_S(t)$, the heat $Q_S(t)$ deviates from its weak-coupling form $Q_w(t)$. Moreover, a significant contribution to the work $W_S(t)$ is observed, even for a time-independent system Hamiltonian, $H_S$. This work is proportional to the magnitude of the shift in $K_S(t)$ and is caused by the time-dependent renormalization of the energy levels of the system energy due to the coupling to the environment. Since $K_S(t)$ is proportional to $\sigma_x$, both work and heat strictly depend only on $\langle \sigma_x\rangle (t)$.

The comparison  of the entropy production rate $\sigma_S(t)$ to its weak coupling form $\sigma_w(t)$ depicted in Fig.\ref{fig:sswe0} shows a very similar behavior of the two quantities. The only difference is a shift of the peak of $\sigma_S(t)$ to earlier times, caused by the different time-dependence of $Q_S(t)$ vs.\  $Q_w(t)$, but this effect is small in the parameter regime considered.

\subsection{Biased ($\varepsilon\neq0$), nonadiabatic ($\Delta/\omega_c<1$) regime }

\begin{figure}[t]
	\includegraphics[width=0.50\textwidth]{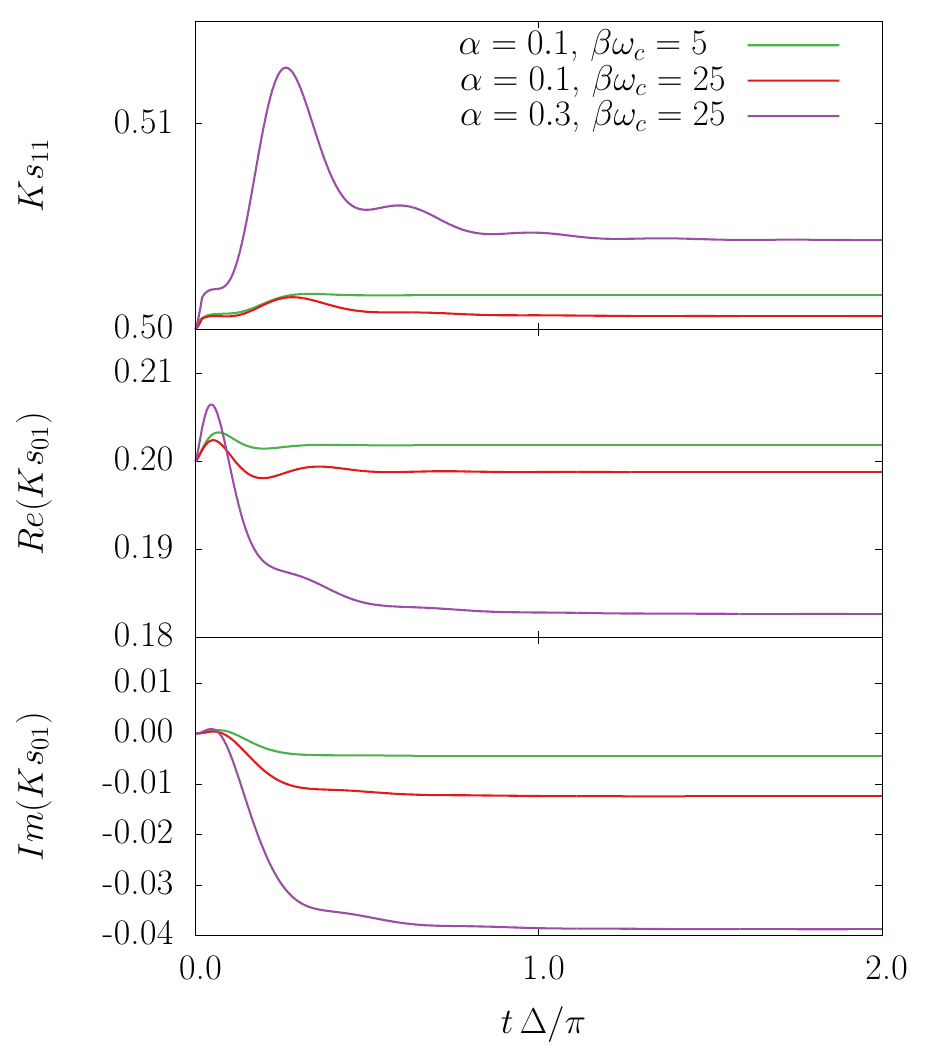}
	\caption{Off-diagonal and diagonal elements of $K_S(t)$ as a function of time for different values of system-bath coupling $\alpha$ and inverse temperature $\beta$. Other parameters are $\varepsilon/\Delta=2.5$, $\Delta/\omega_c=0.2$. Results are given in units of $\omega_c$.
		}
	\label{fig:kse5}
\end{figure}
\begin{figure}[t]
	\includegraphics[width=0.50\textwidth]{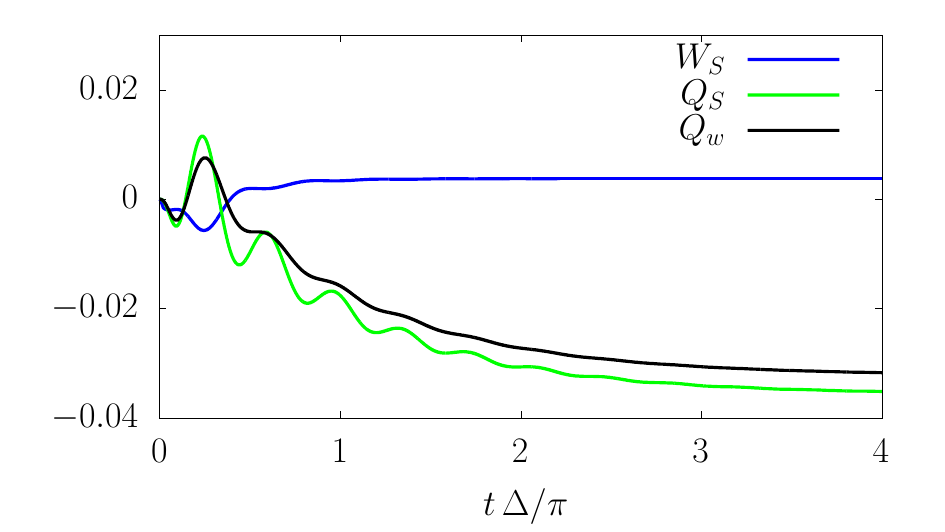}
	\caption{Work $W_S(t)$ and heat $Q_S(t)$ as a function of time, as well as a comparison with the weak-coupling heat $Q_w(t)$. Parameters are $\alpha=0.3$, $\beta \omega_c=25$, $\varepsilon/\Delta=2.5$, $\Delta/\omega_c=0.2$. Results are given in units of $\omega_c$. 
	}
	\label{fig:qwe5}
\end{figure}
\begin{figure}[t]
	\includegraphics[width=0.50\textwidth]{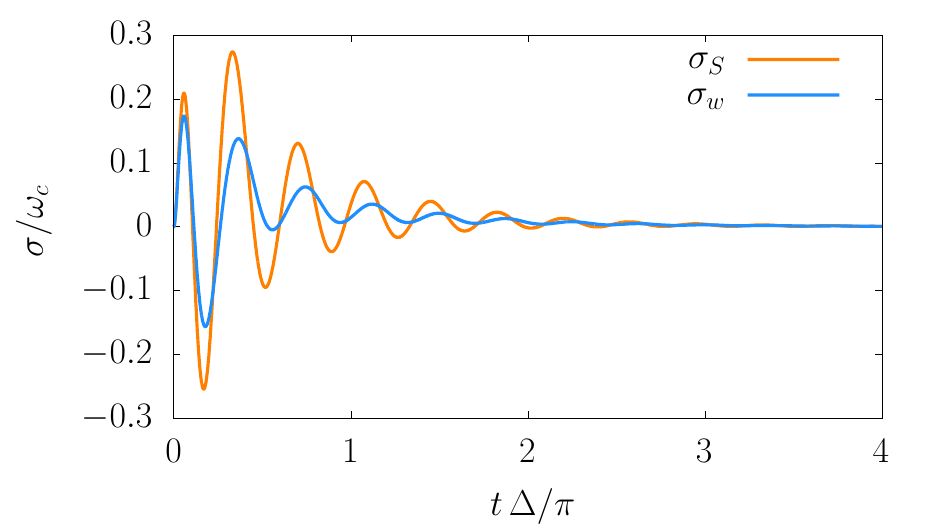}
	\caption{Comparison between the entropy production rate $\sigma_S(t)$ and its weak-coupling version $\sigma_w(t)$. Parameters are $\alpha=0.3$, $\beta \omega_c=25$, $\varepsilon/\Delta=2.5$, $\Delta/\omega_c=0.2$.
	}
	\label{fig:sswe5}
\end{figure}

Next, we discuss results for the biased case in the non-adiabatic regime ($\Delta/\omega_c<1$). The inclusion of a non-zero bias introduces additional complexity to the quantum dynamics and results in an effective Hamiltonian \(K_S(t)\), which assumes a more general form,
\begin{eqnarray} \label{kse5}
	K_S(t) = -(\varepsilon+\delta_{z} (t)) \sigma_z +(\Delta+\delta_{x}(t)) \sigma_x  + \delta_{y}(t) \sigma_y.
\end{eqnarray}  

\noindent
Fig.\ \ref{fig:kse5} shows both the off-diagonal and diagonal elements of $K_S(t)$ in the nonadiabatic regime, $\Delta/\omega_c=0.2$, for different parameters of the system-bath coupling and the temperature.
The off-diagonal elements undergo a similar shift as in the unbiased case, but acquire also an imaginary part, $\delta_{y}(t)$. In addition, and different to the unbiased case, the coupling to the bath induces a time-dependent shift, $\delta_{z}(t)$ of the diagonal elements of $K_S(t)$, which increases slightly with higher temperature. In contrast to the unbiased case, not only the eigenvalues but also the eigenstates of the effective Hamiltonian \(K_S(t)\) are now time-dependent, undergoing a renormalization influenced by the environment.

Fig.\ \ref{fig:qwe5} shows the time evolution of the work $W_S(t)$ and the heat $Q_S(t)$, compared to its weak-coupling form $Q_w(t)$.
From the parameter sets shown in Fig.  \ref{fig:kse5}, we have selected again the most interesting case of larger coupling ($\alpha=1.0$) and low temperature ($\beta \omega_c=25$).  As for the unbiased case, the effects are very similar but less pronounced for smaller coupling and higher temperature.
In contrast to the unbiased case, both $W_S(t)$ and $Q_S(t)$ exhibit pronounced coherent oscillations before they reach a steady state.  
The weak coupling form of the heat, $Q_w(t)$, differs from $Q_S(t)$ both in its less oscillatory character and its smaller absolute value for long times. 

As a consequence, the entropy production rate, depicted in  Fig.\ref{fig:sswe5},  exhibits  in this regime a notable differences among the two definitions. Specifically, the amplitude of $\sigma_S(t)$  is larger than that of  \(\sigma_w(t)\) and there is a phase shift between the two results. The latter is a direct result of the coupling to the thermal bath,
which causes the different time-dependence of $Q_S(t)$ and $Q_w(t)$, which is very pronounced in the parameter regime considered.

\subsection{Biased ($\varepsilon \neq 0$), adiabatic ($\Delta/\omega_c>1$) regime }
\begin{figure}[t]
	\includegraphics[width=0.50\textwidth]{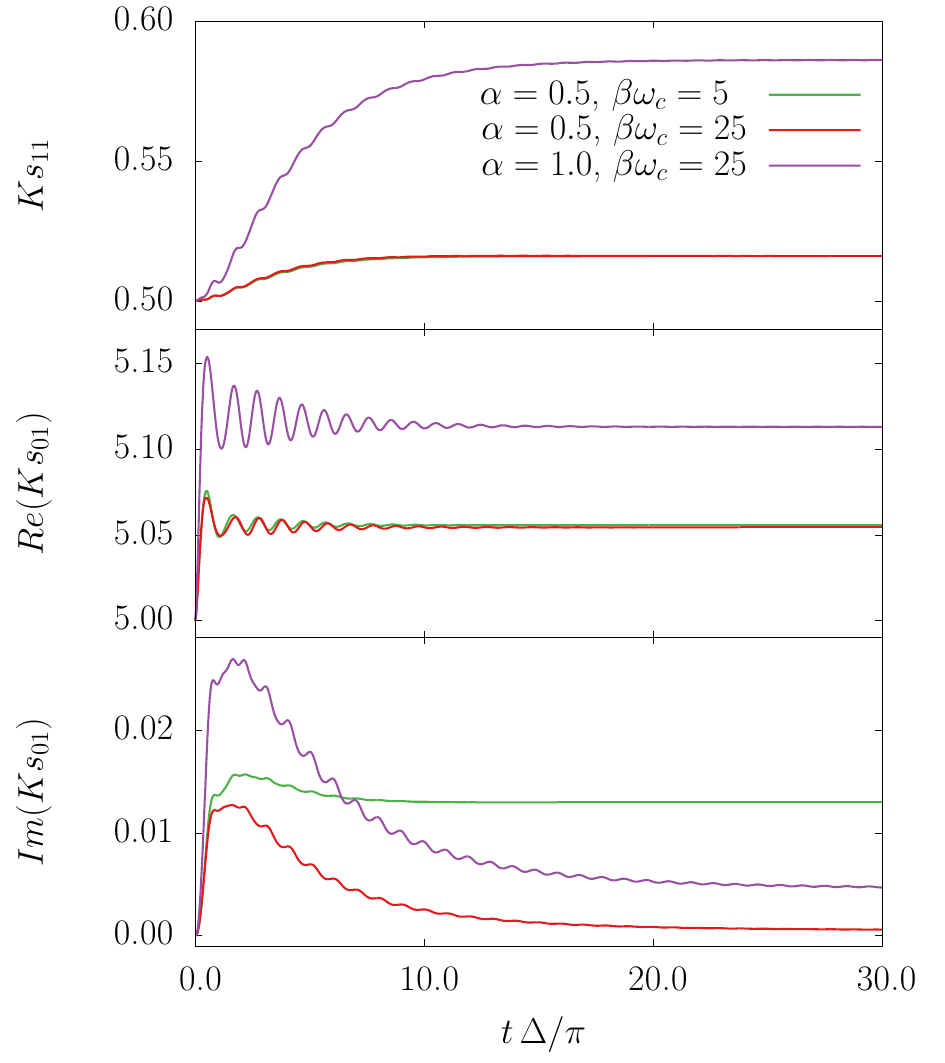}
	\caption{Off-diagonal and diagonal elements of $K_S(t)$ as a function of time for different values of system-bath coupling $\alpha$ and inverse temperature $\beta$. Other parameters are $\varepsilon/\Delta=0.1$, $\Delta/\omega_c=5.0$. Results are given in units of $\omega_c$.
	}
	\label{fig:ksead}
\end{figure}
\begin{figure}[t]
	\includegraphics[width=0.50\textwidth]{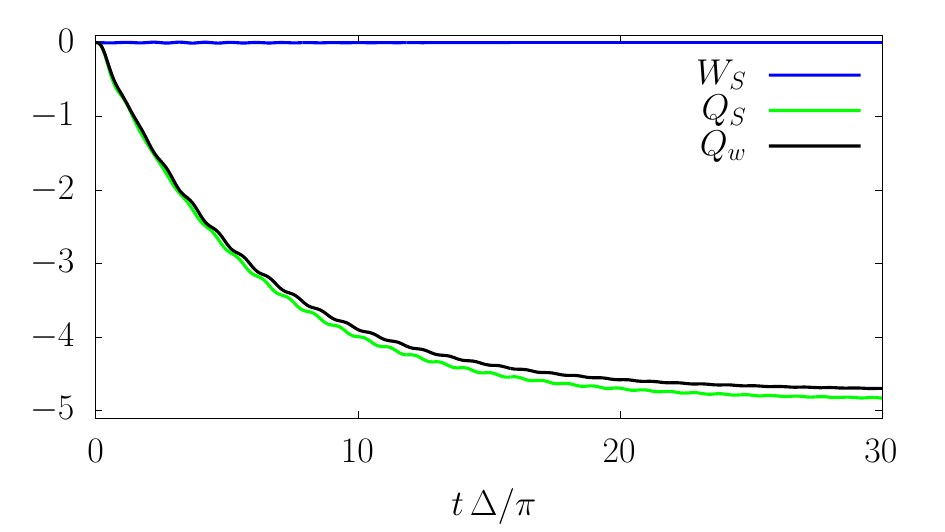}
	\caption{Work $W_S(t)$ and heat $Q_S(t)$ as a function of time, as well as a comparison with the weak-coupling heat $Q_w(t)$. Parameters are $\alpha=1.0$, $\beta \omega_c=25$, $\varepsilon/\Delta=0.1$. Results are given in units of $\omega_c$.		
	}
	\label{fig:qwead}
\end{figure}
\begin{figure}[t]
	\includegraphics[width=0.50\textwidth]{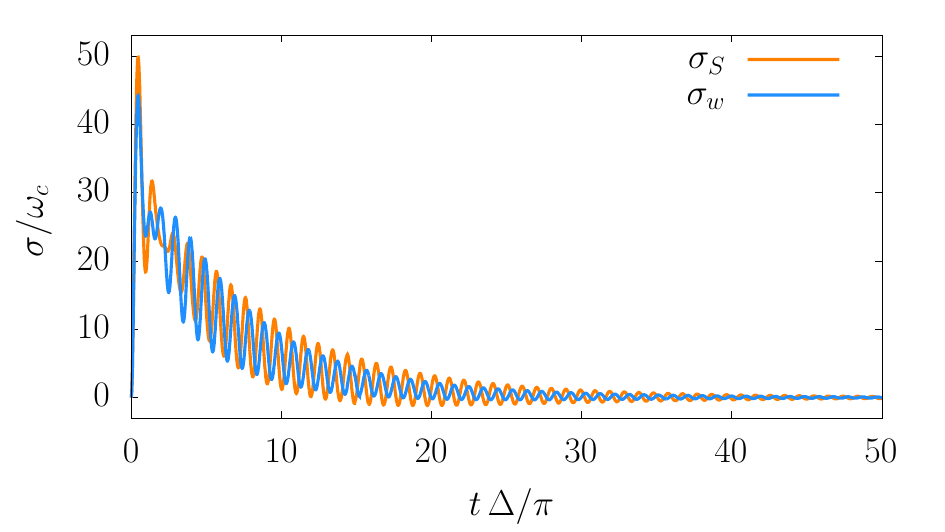}
	\caption{Comparison between the entropy production rate $\sigma_S(t)$ and its weak-coupling version $\sigma_w(t)$. Parameters are $\alpha=1.0$, $\beta \omega_c=25$, $\varepsilon/\Delta=0.1$, $\Delta/\omega_c=5.0$.
	}
	\label{fig:sswead}
\end{figure}

We conclude our discussion with  results in the adiabatic regime $(\Delta / \omega_c > 1)$. In this regime, the spin-boson model exhibits a more complex behavior \cite{Thoss2001}, characterized by oscillatory coherent decay and non-Markovian dynamics \cite{Wenderoth2021} over a larger range of parameters. Here, we discuss results for finite energy bias $\varepsilon \neq 0$. The unbiased adiabatic case shows very similar results to the biased adiabatic case.
The effective Hamiltonian $K_S(t)$ in the biased adiabatic case takes the general form of Eq.\ \eqref{kse5}. 

Fig.\ \ref{fig:ksead} shows the elements of $K_S(t)$ for different parameters. Notably, the time scale over which a steady state is reached is significantly longer than in the non-adiabatic regime. Furthermore, the real part of $K_S(t)$ is less sensitive to the temperature. These two findings are a consequence of the frequency mismatch between system and bath in the adiabatic regime: Since $\Delta>\omega_c$, the energy levels of the system are located in the high-frequency tail of the spectral density $J(\omega)$ (Eq.\ \eqref{eq:debye_spectral_density}), corresponding to a less efficient system-bath coupling and processes which are largely independent on temperature. 

Fig.\ \ref{fig:qwead} shows the time evolution of work $W_S(t)$ and heat $Q_S(t)$, compared to its weak-coupling form $Q_w(t)$. 
In contrast to the nonadiabatic regime, there is little difference between the two forms of the heat and the work $W_S(t)$ is negligible. The small differences in the time dependence of 
$Q_S(t)$ and $Q_w(t)$ results, however, in visible difference between the entropy production rate \(\sigma_S(t)\) and its weak coupling form, depicted in Fig.\ref{fig:sswead}. In particular, a significant phase shift between the two result is apparent.

\section{Conclusions}\label{sec:conclu}

In this paper, we have investigated the thermodynamics of the spin-boson model in the presence of non-Markovian effects and moderate to strong system-bath coupling. Our analysis was based on a novel approach to quantum thermodynamics, which uses the principle of minimal dissipation to obtain a unique decomposition of the generator of time evolution into a Hamiltonian and a dissipative part, allowing to define the internal energy and to identify the contribution of heat and work. To simulate the dynamics of these thermodynamic properties, we employed the HEOM approach, which provides a numerically exact description of the quantum dynamics of the spin-boson model.

The results obtained for a broad range of parameter regimes demonstrate the performance of the approach. In all parameter regimes, the coupling to the environment results in a time-dependent effective Hamiltonian, even if the original system Hamiltonian is time-independent. This time-dependence, which is particularly pronouned for stronger coupling, reflects the work done by the environment on the system without additional external work protocols. 
This suggest that it may be possible to engineer environments in order to enhance work extraction from a quantum system \cite{Monsel2020,Manzano2018}. The analysis of thermodynamic properties such as heat and the entropy production shows that the results obtained within the principle of minimal dissipation differ significantly from those obtained using traditional expressions defined in the weak coupling limit. This difference is especially noticeable in the nonadiabatic regime of the spin-boson model.
The observed influence of strong-system bath coupling on entropy production may also have a non-trivial impact on thermodynamic uncertainty relations \cite{Horowitz2016,Popescu2014}.

The spin-boson model considered in this paper relaxes for long times to an equilibrium steady state. The extension of the quantum thermodynamics approach used here to situations with nonequilibrium steady states, such as, e.g., the nonequilbrium spin-boson model \cite{VELIZHANIN2008} or nanojunctions \cite{ThossEvers2018}, will be an interesting topic for future work.

\acknowledgments

This project was supported by the European Union's Framework Programme 
for Research and Innovation Horizon 2020 (2014-2020) under the 
Marie Sk\l{}odowska-Curie Grant Agreement No.~847471. Furthermore, support by the state of Baden-Württemberg through
bwHPC and the DFG through Grant No. INST 40/575-1
FUGG (JUSTUS 2 cluster) is gratefully acknowledged.

\appendix

\section{Truncation of the HEOM hierarchy}\label{app-A}

\begin{figure}[t]
	\includegraphics[width=0.50\textwidth]{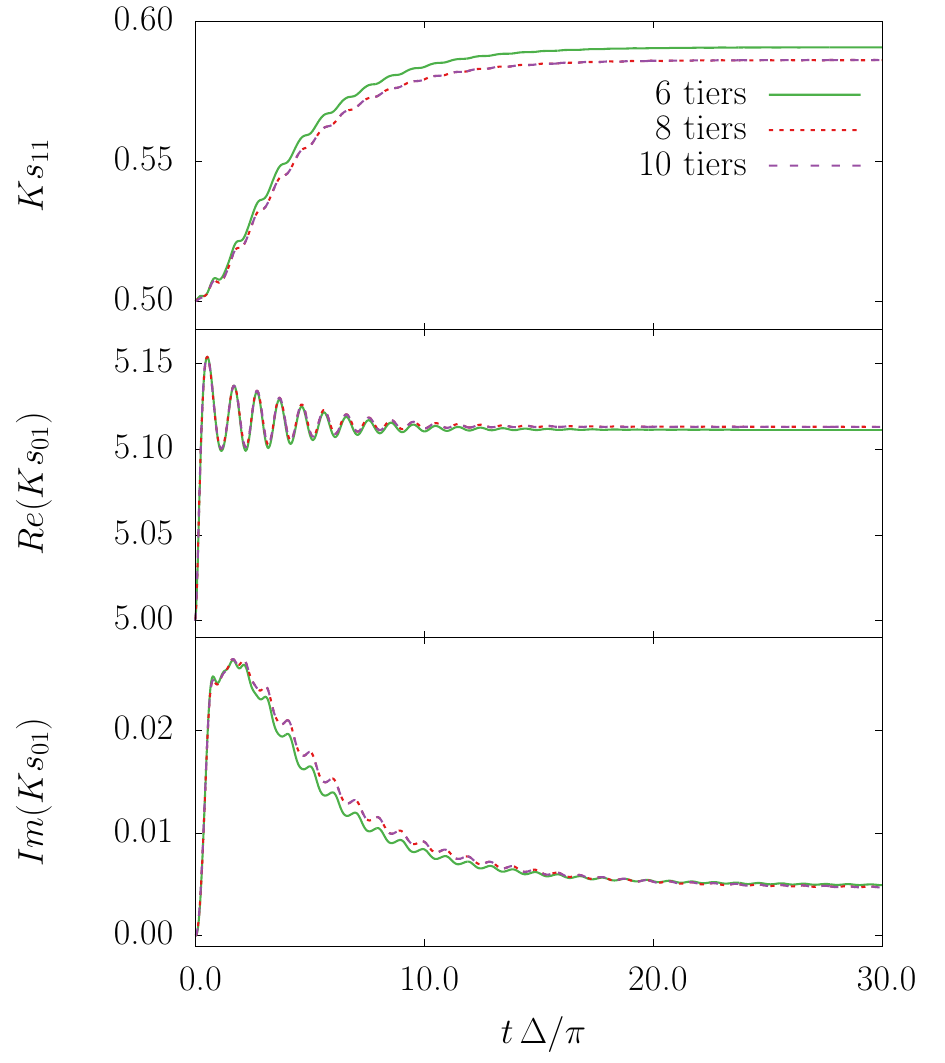}
	\caption{Convergence of off-diagonal and diagonal elements of $K_S(t)$. Parameters are $\alpha=1.0$, $\beta \omega_c=25$,$\varepsilon/\Delta=0.1$, $\Delta/\omega_c=5.0$. Results are given in units of $\omega_c$.
	}
	\label{fig:tier}
\end{figure}

Fig. \ref{fig:tier} illustrates the convergence of the off-diagonal and diagonal elements of $K_S(t)$ with respect to the truncation of the hierarchy for the spin-boson model, using as an example the data in Fig.\ \ref{fig:ksead} for the case $\alpha=1.0$, $\beta \omega_c=25$. Both diagonal and off-diagonal elements of the effective Hamiltonian $K_S(t)$ are converged for a truncation of the hierarchy after 8 tiers, whereas all results in this work were calculated using 10 tiers. 

\section{Derivation of the pseudo-Kraus representation in Eq. \eqref{eq:krausform} }\label{app-B}

Here, we provide some details on the derivation of the pseudo-Kraus representation in Eq. \eqref{eq:krausform}. We assume a factorized initial state for the composite system $S+B$, i.e. $\rho_{S+B}(0)=\rho_{S}(0)\otimes\rho_{B}(0)$. It is then possible to write a time-local general master equation, governing the density matrix $\rho_S(t)$ of the open system $S$,
\begin{equation} \label{tcl-meq2}
	\frac{d}{dt}\rho_S(t) = \Lt_t\rho_S(t) \, .
\end{equation}

Let $\{ \tau_a = \ket{j_a}\bra{k_a}, a=1,2,...,N^2\}$ denote an orthonormal basis for the operators in the Hilbert space of system $S$, with $j_a,k_a=1,2,\ldots,N$. The $N^2\times N^2$ Choi matrix $M(t)$ is defined as
\begin{equation}
	M_{a b}(t) := \bra{j_a}\, \Big[\Lt_t \ket{k_a} \bra{j_b}\Big] \ket{k_b} \, .
	\label{eq:Choim}
\end{equation}

Using the Choi matrix, the action of the generator $\Lt_t$ can be written as
\begin{eqnarray}
	\Lt_t \rho_S(t) &=& \sum_{ab} M_{ab}(t) \tau_a \, \rho_S(t)\, \tau_b^\dagger  \label{choirep} .
\end{eqnarray}
Since $\Lt_t$ maps Hermitian operators to Hermitian operators, $M(t)$ is Hermitian, and the eigenvalues and normalized eigenvectors of $M(t)$ are given by 
\begin{equation} \label{eig1}
	M(t) \vec{v}(k) = \theta_k(t) \vec{v}(k),
\end{equation}
where $\vec{v}(k) \in \mathbb{C}^{N^2}$.
Thus, one can always decompose $M(t)$ as 
\begin{equation} \label{sdecomp}
	M(t) = \sum_k \theta_k(t) \vec{v}(k) \,\vec{v}(k)^\dag \, ,
\end{equation} 
and from Eqs.~(\ref{choirep}) and (\ref{sdecomp}) it follows
\begin{align}
	\Lt_t \rho_S(t) &= \sum_{k,a,b} \theta_k(t) \, v_a(k) \, v^*_b(k) \, \tau_a \, \rho_S(t)\, \tau_b^\dagger \nonumber \\
    &=\sum_{k,a,b} \theta_k(t) \, v_a(k) \, \tau_a \,\rho_S(t)\, \tau_b^\dagger \, v^*_b(k) \, \nonumber \\
	&= \theta_k(t) E_k(t) \rho_S(t) E_k^{\dagger}(t) \,,
\end{align}
This corresponds to Eq.~(\ref{eq:krausform}), where the Kraus operators are given by
\begin{equation} \label{krausdef}
	E_k(t) := \sum_{a} \theta_k(t) \, v_a(k) \, \tau_a .
\end{equation}

In our approach to evaluate the pseudo-Kraus representation for $\Lt_t$, we first determine a representation of the dynamical map $\Phi_t$ by solving Eq.~\eqref{eq:general_HQME} for different orthogonal initial states of the system. The generator of the dynamical map is then obtained as $\Lt_t=(\partial_t{\Phi}_t){\Phi_t}^{-1}$, which provides the Choi matrix via Eq. \eqref{eq:Choim}.


\bibliography{biblio_nonmarkov}

\end{document}